\documentclass[aps,twocolumn,showpacs,superscriptaddress,floatfix,nofootinbib]{revtex4-1}
\usepackage{amsfonts}

\usepackage{graphicx}
\usepackage{amsmath}
\usepackage{amssymb}
\usepackage{bm}
\usepackage{gensymb}

\usepackage{hyperref}

\begin{document}
\title{Emergent Haldane phase in the $S=1$ bilinear-biquadratic Heisenberg model on the  square lattice }

\author{Ido Niesen}
\affiliation{Institute for Theoretical Physics and Delta Institute for Theoretical Physics, University of Amsterdam, Science Park 904, 1098 XH Amsterdam, The Netherlands}

\author{Philippe Corboz}
\affiliation{Institute for Theoretical Physics and Delta Institute for Theoretical Physics, University of Amsterdam, Science Park 904, 1098 XH Amsterdam, The Netherlands}

\date{May 2, 2017}

\begin{abstract}
Infinite projected entangled pair states simulations of the $S=1$ bilinear-biquadratic Heisenberg model on the square lattice reveal an emergent Haldane phase in between the previously predicted antiferromagnetic and 3-sublattice 120$^\circ$ magnetically ordered phases.  This intermediate phase preserves SU(2) spin and translational symmetry but breaks lattice rotational symmetry, and it can be adiabatically connected to the Haldane phase of decoupled $S=1$ chains. Our results contradict previous studies which found a direct transition between the two magnetically ordered states. 
\end{abstract}



\maketitle


The search for novel states of matter in quantum many-body systems is one of the most active areas in condensed matter physics. A fascinating example is the ground state of the $S=1$ antiferromagnetic Heisenberg chain which, unlike the $S=1/2$ chain, exhibits an energy gap, exponentially decaying spin-spin correlations, and gapless edge excitations in case of open boundaries. Thanks to Haldane's pioneering work and conjecture that such a gapped state emerges in integer Heisenberg spin chains in general~\cite{haldane83,haldane83b}, this phase has been named after him.

The Haldane phase also extends  to related $S=1$ models, such as weakly-coupled $S=1$ Heisenberg chains~\cite{sakai89,koga00b,kim00,matsumoto01,wierschem14}  which are realized in several quasi-1D materials~\cite{meyer82,renard87,monfort96,honda97,honda98,uchiyama99}, or the $S=1$ bilinear-biquadratic Heisenberg (BBH) chain with Hamiltonian
\begin{equation*}
{\cal H} = \sum_{\langle i,j \rangle} \left[ \cos(\theta)  {\bf S_i \cdot S_j} + \sin(\theta)  \left({\bf S_i \cdot S_j}\right)^2 \right],
\end{equation*}
for $\theta$ in between $-\pi/4$ and $\pi/4$. More recently, the Haldane phase has been understood as a simple example of a symmetry protected topological (SPT) phase~\cite{gu09, pollmann10, pollmann12,chen13b}.

In the present work we focus on the BBH model in two dimensions, which has gained much interest in recent years~\cite{papanicolaou88,harada02,tsunetsugu06,lauchli06,tsunetsugu07, bhattacharjee06,toth12,bieri12,oitmaa13}; firstly, due to its possible connection to the triangular lattice compounds NiGa$_2$S$_4$~\cite{nakatsuji05} and Ba$_3$NiSb$_2$O$_9$~\cite{COBA,cheng11,bieri12,fak17}, and secondly, for $\theta=\pi/4$, the model is equivalent to the SU(3) Heisenberg model which can be experimentally realized using ultra-cold fermionic atoms in optical lattices~\cite{wu2003,honerkamp2004,cazalilla2009,gorshkov2010}.  The latter has been shown to exhibit 3-sublattice order on the square and triangular lattices~\cite{toth2010,Bauer12}, and an important question concerns the stability of this phase away from the SU(3) symmetric point. Previous studies on the square lattice based on linear flavor-wave theory~\cite{toth12}, exact diagonalization~\cite{toth12}, and series expansion~\cite{oitmaa13}  predicted a direct transition between the AF and the 3-sublattice phase for $\theta\approx 0.2\pi$. However, the accurate study of this parameter regime remains very challenging because Quantum Monte Carlo suffers from the negative sign problem. 

\begin{figure}
\includegraphics[width=1\columnwidth]{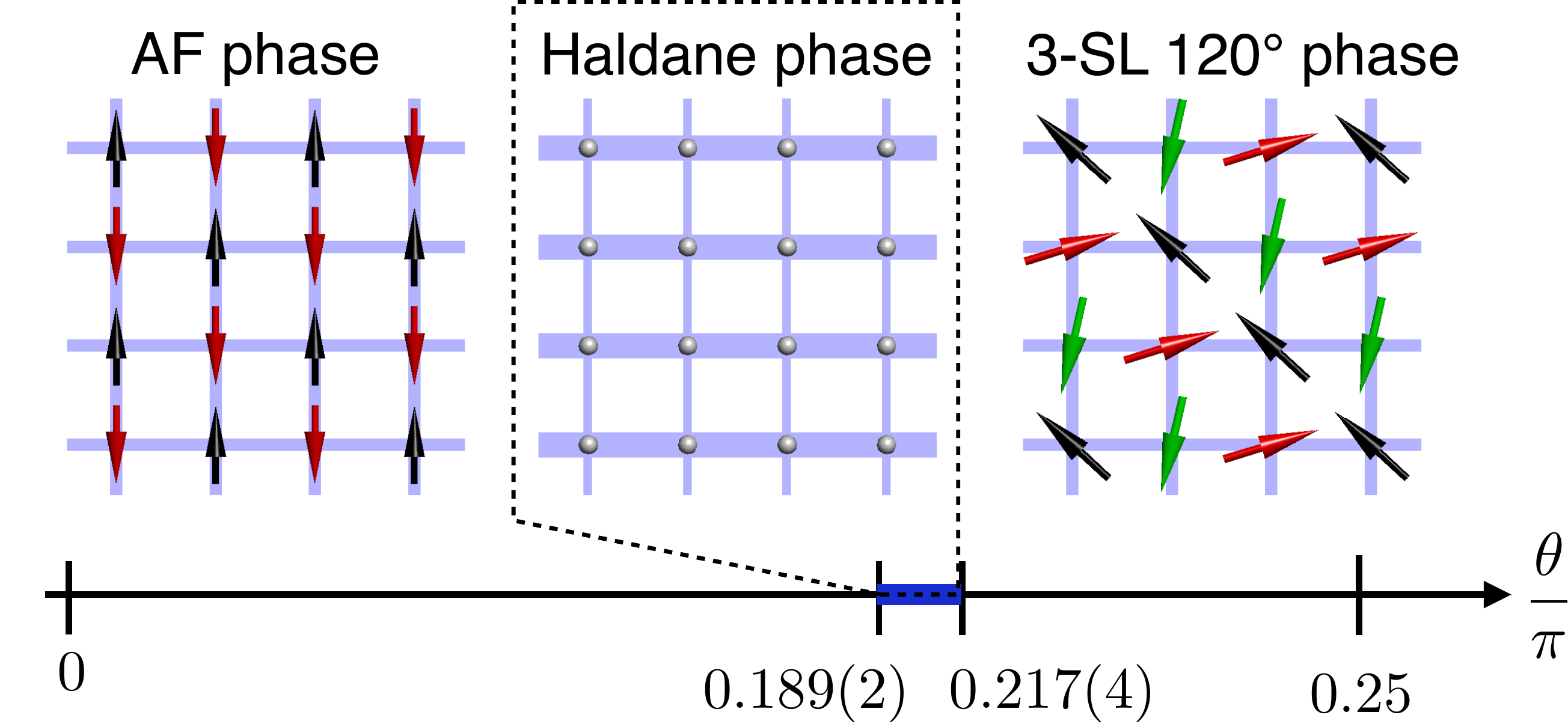}
\caption{Haldane phase emerging in between the antiferromagnetic (AF) and the 3-sublattice 120$^{\circ}$ ordered phases.}
\label{fig:pd}
\end{figure}

 
In this paper we show, using state-of-the-art tensor network simulations, that in between the AF and the 3-sublattice phase an intermediate quantum paramagnetic phase emerges which preserves translational and SU(2) spin symmetry, but breaks lattice rotational symmetry (see Fig.~\ref{fig:pd}). We identify this intermediate phase as the Haldane phase by showing that it can be adiabatically connected to the Haldane phase of decoupled $S=1$ chains. This result at first appears surprising in view of the fact that for $\theta=0$  already a small interchain coupling $J_y > J^c_y = 0.0436$~\cite{matsumoto01} is sufficient to destabilize the Haldane phase. However, we show that with increasing $\theta$ the critical interchain coupling $J^c_y(\theta)$ separating the Haldane phase from the AF phase dramatically increases, and eventually reaches the isotropic 2D limit.

{\emph{Method  --} Our results have been obtained using infinite projected entangled-pair states~\cite{verstraete2004,jordan2008,corboz2010} (iPEPS, also called tensor product state~\cite{nishino01,nishio2004}) --- a variational tensor-network ansatz to efficiently represent ground states in two dimensions directly in the thermodynamic limit. It can be seen as a natural generalization of matrix product states (the underlying ansatz of the powerful density-matrix renormalization group method~\cite{white1992}) to two dimensions. This approach has already been applied successfully to a variety of challenging problems in the field of frustrated magnetism and strongly correlated electrons, see e.g. Refs.~\cite{corboz11-su4,wang11_j1j2, Zhao12,  Corboz12_su4, xie14, Corboz13_shastry, gu2013, matsuda13, Osorio14, corboz14_tJ, corboz14_shastry, picot15a, picot15,corboz16,nataf16,liao16,zheng16}  and references therein.

The iPEPS ansatz consists of a network of order-5 tensors on a square lattice, with one tensor per lattice site. Each tensor has a physical index carrying the local Hilbert space of a site and four auxiliary indices which connect to the four nearest-neighboring tensors. Each auxiliary index goes over $D$ elements, called the bond dimension, which controls the accuracy of the ansatz. For translational invariant states an ansatz with the same tensor on each lattice site can  be used, however, if translational symmetry in the ground state is broken, a larger unit cell of tensors is required. For example, to reproduce an antiferromagnetic state two different tensors (one for each sublattice) is needed, whereas the 3-sublattice 120$^{\circ}$  ordered state shown in Fig.~\ref{fig:pd} requires a unit cell with 3 different tensors. In practice we run simulations using different unit cells to find out which structure yields the lowest energy state. 

For more details on the method we refer to Refs.~\cite{corboz2010,Corboz13_shastry,Phien15}. For the experts we note that the optimization of the tensors has been done via an imaginary time evolution with the full update optimization \cite{corboz2010} (or fast-full update~\cite{Phien15}), except for the simulations of the anisotropic model where we used the computationally cheaper simple-update optimization~\cite{vidal2003-1,jiang2008}. The contraction of the infinite tensor network is done by a variant~\cite{corboz2011,corboz14_tJ} of  the corner-transfer matrix method~\cite{nishino1996,orus2009-1}. We also exploited the U(1) symmetry~\cite{singh2010,bauer2011} to increase the efficiency (except in the 3-sublattice phase which breaks U(1) symmetry).


{\emph{AF and 3-sublattice phases --} 
We first discuss the well-established limits of the phase diagram in the range $\theta\in[0,\pi/4]$. For $\theta=0$ the model reduces to the $S=1$ Heisenberg model where the ground state exhibits antiferromagnetic order. Unlike for $\theta>0$, Quantum Monte Carlo has no negative sign problem in this limit, and the sublattice magnetization $m=\sqrt{ \langle S_x\rangle^2 + \langle S_y\rangle^2  + \langle S_z\rangle^2}$ in the thermodynamic limit has been accurately determined: $m=0.805(2)$~\cite{matsumoto01}. Our iPEPS result extrapolated to the infinite $D$ limit (see Fig.~\ref{fig:mex}), $m=0.802(7)$, is in agreement with this value. For $\theta=\pi/4$ the model is equivalent to the $SU(3)$ Heisenberg model (with the fundamental representation on each site), for which a 3-sublattice ordered state has been predicted by several methods~\cite{toth2010,Bauer12}, including previous iPEPS simulations. For $\theta$ slightly below $\pi/4$ this order corresponds to a $120^\circ$ order formed by the spins on 3 sublattices (see  Fig.~\ref{fig:pd}).

\begin{figure}
\includegraphics[width=1\columnwidth]{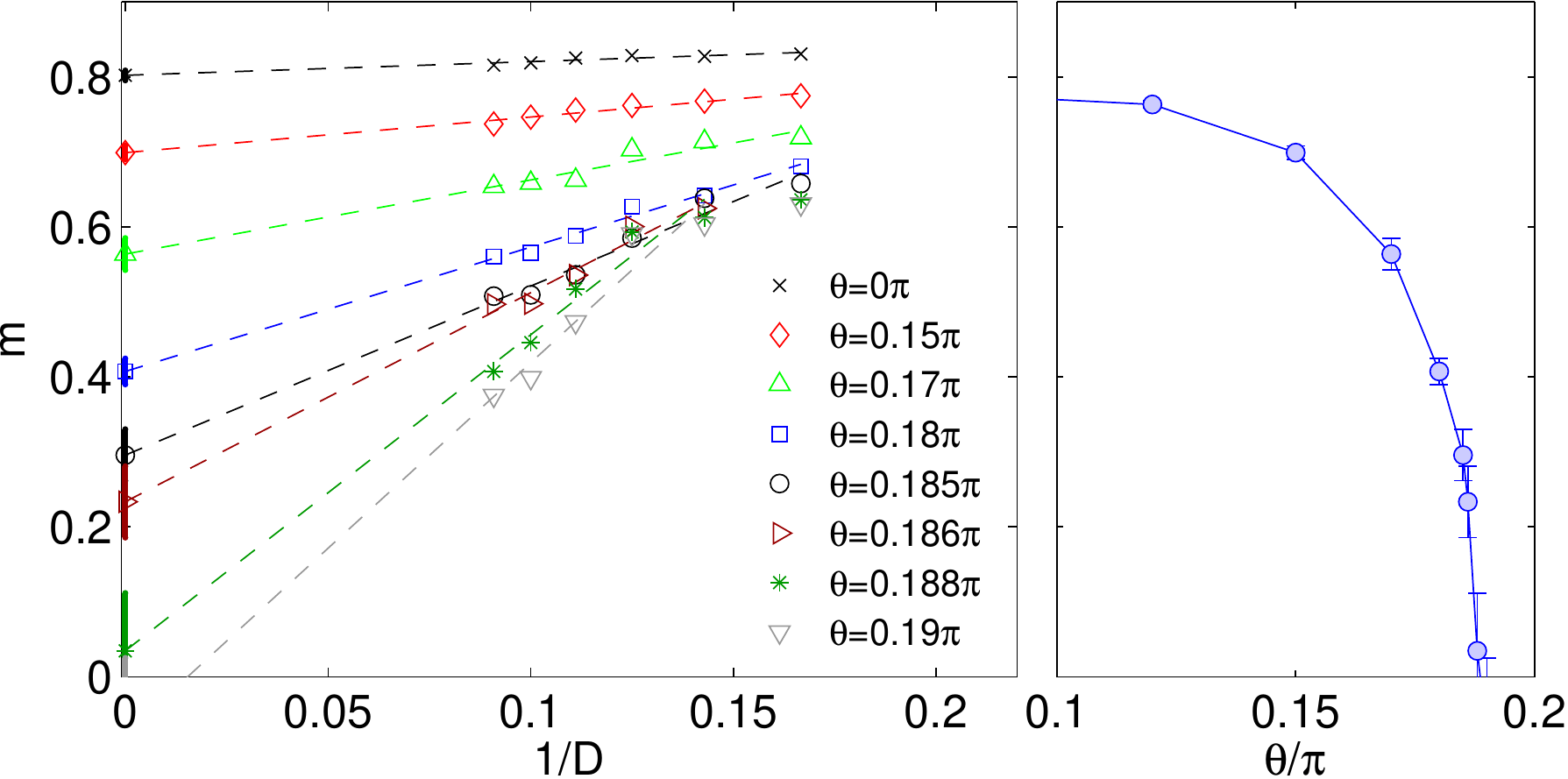}
\caption{(a) iPEPS results (full update) for the local magnetic moment $m$ as a function of inverse bond dimension for different values of $\theta$. (b) Extrapolated values of $m$ as a function of~$\theta$.} 
\label{fig:mex}
\end{figure}

{\emph{Intermediate phase --} 
In Refs.~\cite{toth12,oitmaa13} a direct transition between the AF state and the 3-sublattice state has been predicted to occur around $\theta\approx0.2\pi$ based on exact diagonalization, linear flavor-wave theory and series expansion. We first attempt to reproduce this result with iPEPS by determining the critical value $\theta_c$ for which the energies of the two states --- distinguished by different unit cells --- intersect. To do so, we initialize a simuluation from deep within the AF (3-sublattice) phase, and slowly increase (decrease) $\theta$ up to the point where the energies cross. The resulting critical value $\theta_c \approx 0.21\pi$ (for $D=10$) is close to the previous prediction. However, from a systematic analysis of the AF order parameter shown in Fig.~\ref{fig:mex} we find that the AF order actually \emph{vanishes long before $\theta_c$}, i.e. that the AF phase is only stable up to $\theta=0.189(2)\pi$ \cite{COITMAA}. This indicates the presence of an intermediate non-magnetic phase in between the AF- and the 3-sublattice phase.

We next explore the region around $\theta=0.2\pi$ in more detail. When starting from different random initial states with a 2-sublattice ansatz, we observe a competition between a weakly magnetized state and a non-magnetized state which breaks lattice rotational symmetry but preserves SU(2) and translational symmetry \cite{CVBC}. This non-magnetized state can also be found by restricting the simulation to a 1-site unit cell~\cite{CU}; it exhibits the lowest variational energy for large $D$.

The rotational symmetry breaking manifests itself in energy differences in $x$ and $y-$direction, as illustrated by the different thicknesses of the bonds in the middle panel of Fig.~\ref{fig:pd}, reminiscent of coupled 1D chains. Since the ground state of the BBH chain for $\theta \in [0,\pi/4]$ is in the Haldane phase, the question naturally arises whether the intermediate 2D phase could possibly be adiabatically connected to the Haldane phase by continuously decreasing the $y$-coupling to zero. A first hint that this picture is correct comes from the observation that when initializing the iPEPS as a product of chains in the 1D Haldane phase the simulation converges to the same non-magnetized state as with randomly initialized tensors. In order to confirm this picture we study the stability of the Haldane phase in the anisotropic BBH model in the following.

\begin{figure}
\includegraphics[width=1\columnwidth]{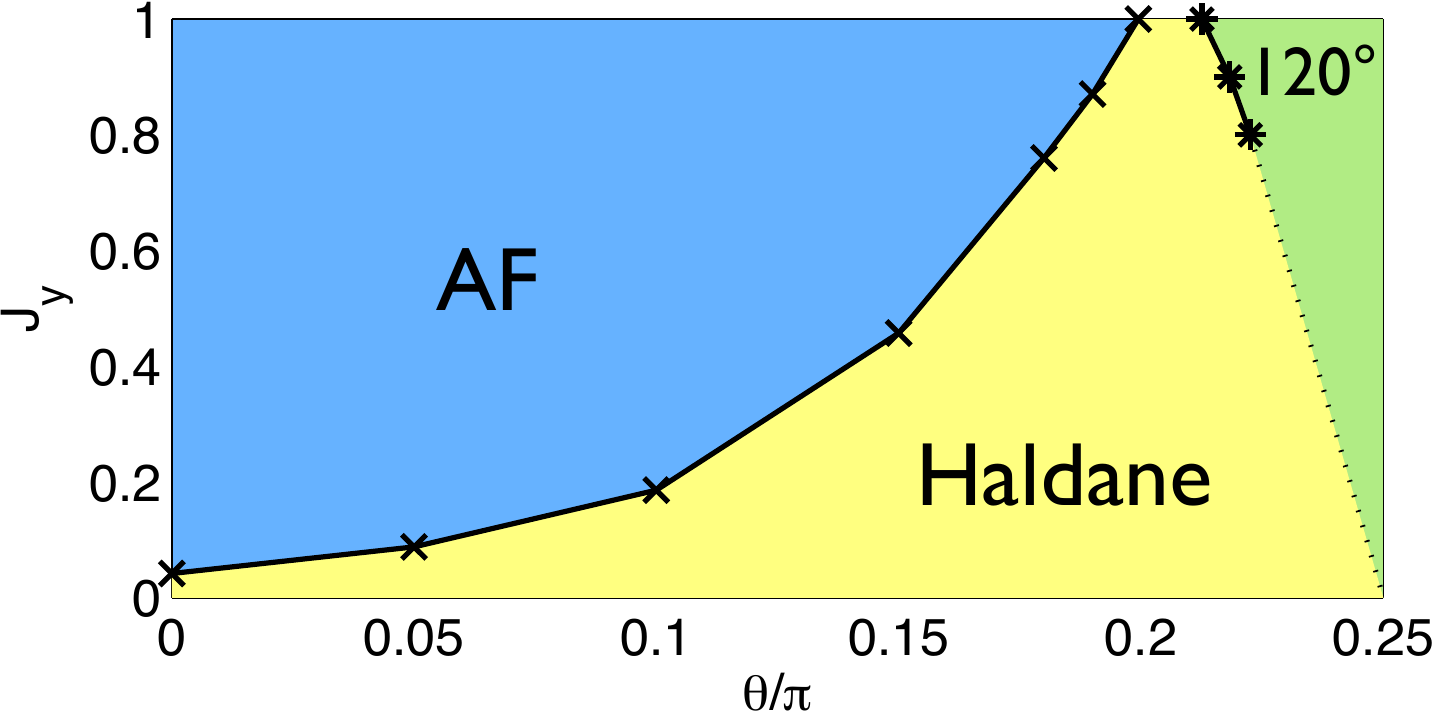}
\caption{The phase diagram of the anisotropic bilinear-biquadratic $S=1$ model. The phase boundaries were estimated based on $D=10$ simple update simulations, see supplemental material~\cite{SUPPL}.}
\label{fig:aniso}
\end{figure}

{\emph{Anisotropic model --}
We introduce different coupling strengths in $x$ and $y-$direction and study the phase diagram in the $(\theta, J_y)$ plane (setting $J_x=1$ in the following). For $J_y=0$ the model simply reduces to independent $S=1$ chains, which are known to lie in the Haldane phase (for $\theta\in[0,\pi/4[$). The goal is now to determine the critical coupling $J_y^c(\theta)$ separating the Haldane phase from the AF phase (or the 3-sublattice phase for large~$\theta$), for different values of $\theta$. In order to obtain an estimate of the phase transition for a fixed value of $\theta$ we initialize the iPEPS in the Haldane phase and in the AF phase (or 3-sublattice phase) respectively, run simulations for different values of $J_y$, and determine the value $J_y^c(\theta)$ for which the energies of the two states intersect, using a fixed bond dimension $D=10$ and the simple update optimization (see data in the supplemental material~\cite{SUPPL}). We note that this approach  provides only an approximate phase boundary, in contrast to the extrapolated full update simulations used in the isotropic case. However, it is computationally much more efficient, which becomes significant when probing the extended two-dimensional parameter space $(\theta, J_y)$. Moreover, a comparison with Monte Carlo and full update results (see below) indicates that the relative error on the phase boundary is only a few percent, which is accurate enough for our current purpose.

For $\theta=0$ we find a critical value $J_y^c(0)=0.042$ which is close to the Quantum Monte Carlo result 0.0436 from Ref.~\cite{matsumoto01} (see also Refs.~\cite{sakai89,koga00b,kim00, wierschem14}). This value lies far away from the isotropic 2D limit $J_y=1$. However, we find that $J_y^c(\theta)$  monotonously increases  with  $\theta$ as shown in the phase diagram in Fig.~\ref{fig:aniso}, and that beyond $\theta_c = 0.200\pi$ no phase transition occurs, showing that the 1D Haldane phase can indeed be adiabatically connected to the isotropic 2D limit. Finally, for $0.213\pi \leq \theta < \pi/4 $ we find a finite transition value $J_y^c(\theta)$ between the Haldane and the 3-sublattice phase which decreases with increasing $\theta$. 

Comparing to the full update results (Fig.~\ref{fig:mex}), which predict the two transitions to be at $0.189(2)\pi$ and $0.217(4)\pi$ respectively, we see that the simple update underestimates the extent of the Haldane phase at the isotropic point. Moreover, it does so by a margin of at most $0.01\pi$ (and by much less for $\theta=0$), indicating that the continuous path that connects the intermediate 2D phase to the 1D Haldane phase persists also when taking the error margin on the phase boundary into account.


{\emph{Transition from Haldane to 3-sublattice phase --}  
We next focus again on the isotropic 2D case ($J_y = 1$) and accurately determine the  transition from the Haldane to the 3-sublattice phase by pushing the simulations up to $D=16$ (Haldane state)  and $D=10$ (3-sublattice state) using the full update optimization, and  compare the energies of the two states in the infinite $D$ limit. Figure \ref{fig:eex}(a) shows the energies extrapolated in the so-called truncation error $w$ (see Ref.~\cite{corboz16} for details). For $\theta=0.21\pi$ the state in the Haldane phase is clearly lower than the 3-sublattice state, whereas for $\theta=0.22\pi$ the opposite is true. By linear interpolation of the energies, taking into account the extrapolation error, we find a critical value of $\theta_c=0.217(4)\pi$. 
Finally, the squares in Fig.~\ref{fig:eex}(b) show the difference in bond energies $\Delta E = E_y - E_x$ in $x$ and $y-$direction of the Haldane state. In the infinite $D$ limit $\Delta E$ tends to a finite value, e.g. $\Delta E=0.07(1)$ for $\theta=0.21\pi$, which shows that the rotational symmetry in the Haldane phase is indeed spontaneously broken.

\emph{Nature of the phase transitions in isotropic case --}
Because the Haldane and 3-sublattice phases break different translational and rotational symmetries, the corresponding phase transition is expected to be first order. This picture is confirmed by the occurrence of hysteresis around the transition point, which allows us to simulate both phases on both sides of the phase transition. Moreover, the sublattice magnetization is strictly positive throughout the 3-sublattice phase (see Fig.~\ref{fig:eex}(b))  |  even for $\theta=0.21\pi$ where the 3-sublattice state is no longer the lowest energy state | implying that the magnetization does \emph{not} go to zero when approaching the transition from above. Since the magnetization is zero in the Haldane phase, it jumps to zero at the transition, showing that the transition is clearly of first order.

As for the AF to Haldane phase transition, the absence of a clear hysteresis in the full update simulations and the fact that the sublattice magnetization in the AF phase goes to zero as we approach the Haldane phase indicates either an unconventional second order or a weak first order phase transition. However, due to the error bars in Fig.~\ref{fig:mex}(b) close to the critical point, we cannot exclude one of the two based on our data.

\begin{figure}
\includegraphics[width=1\columnwidth]{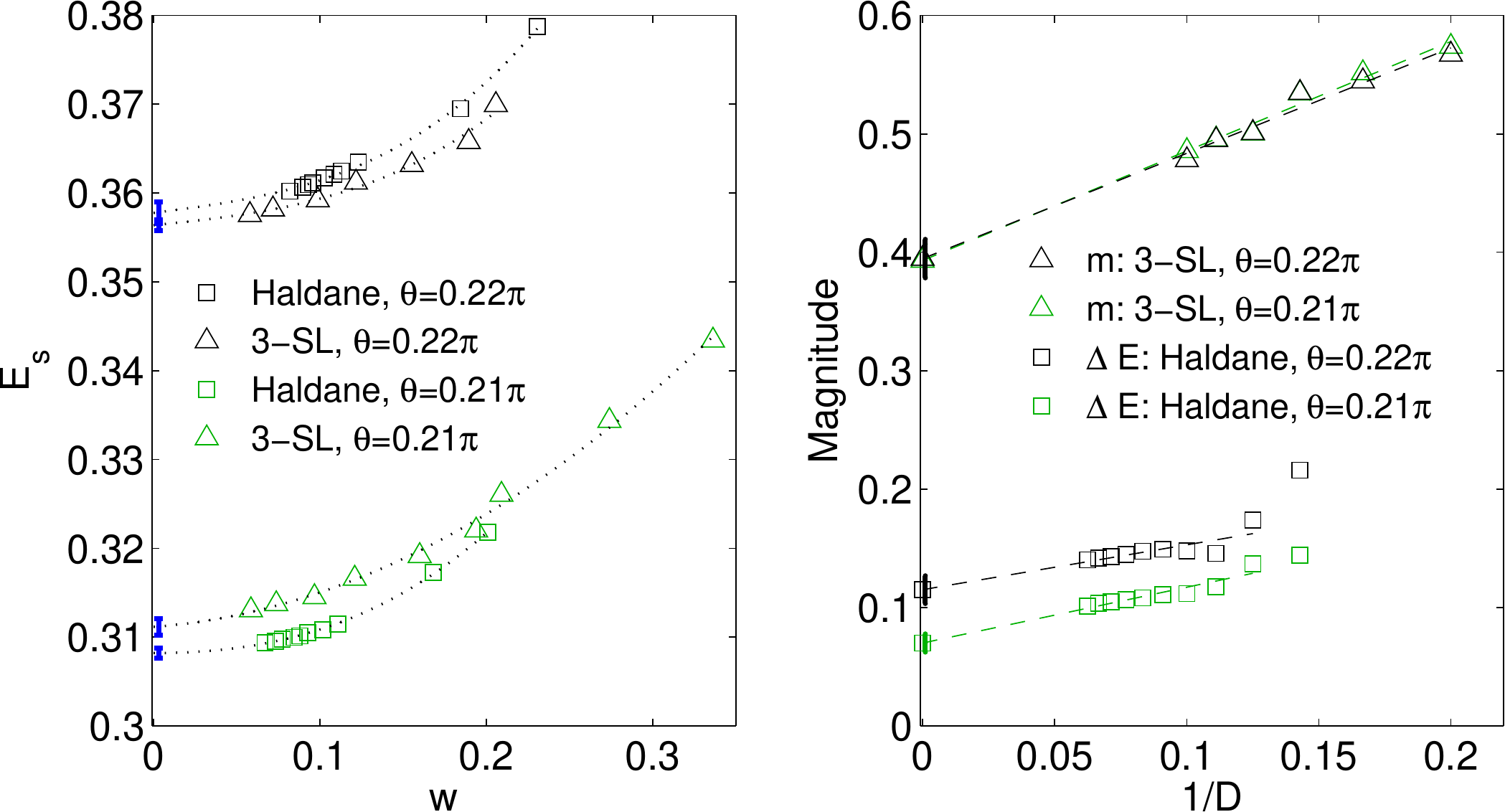}
\caption{(a) Energy per site (full update) of the Haldane and 3-sublattice (3-SL) state for two different values of $\theta$, plotted as a function of the truncation error $w$. (b) Local magnetic moment~$m$ (triangles) of the 3-sublattice state, and the difference in bond energies $\Delta E$ in x and y-direction in the Haldane state (squares), plotted as a function of $1/D$. Note that $m$ and $\Delta E$ are zero in the Haldane and 3-sublattice states respectively.}
\label{fig:eex}
\end{figure} 

 
{\emph{Conclusion --}}
We have studied the $S=1$ BBH model on a square lattice where, in contrast to previous predictions, we found an intermediate quantum paramagnetic  phase between the AF and 3-sublattice 120$^\circ$ magnetically ordered phases in the parameter range $\theta/\pi\in[0.189(2), 0.217(4)]$. This intermediate phase is characterized by (1) translational symmetry, (2)~an absence of magnetic and quadrupolar order, i.e. SU(2) spin symmetry is preserved, and (3) a spontaneous rotational symmetry breaking with stronger bonds in $x$ (or $y$) direction (i.e. lattice nematic order). The above features are reminiscent of the ones of weakly-coupled $S=1$ chains in the Haldane phase, which motivated us to study the anisotropic BBH model. With increasing $\theta$ we found that the critical coupling $J^c_y(\theta)$ separating the Haldane and AF phases monotonically increases up to $\theta = 0.189(2)\pi$, after which no phase transition occurs as a function of $J_y$, i.e. the Haldane phase persists all the way up to the isotropic 2D limit. From this we identified the intermediate phase as a continuous 2D extension of the Haldane phase. 

It is interesting to note that a similar situation has previously been encountered in the $J_1$-$J_2$ $S=1$ Heisenberg model on a square lattice~\cite{jiang09}, in which an intermediate Haldane phase between an AF and a stripe phase appears that also survives up to the isotropic limit. Moreover, our findings provide an additional example of a nematic quantum paramagnet which in Ref.~\cite{wang15} was proposed to likely emerge in spin-1 systems with competing interactions and suggested to be potentially relevant to understand the nematic phase in the iron-based superconductor FeSe.

Finally, our results further highlight the potential of iPEPS as a powerful tool for challenging open problems in frustrated magnetism where Quantum Monte Carlo suffers from the negative sign problem. As a future work it will be interesting to see whether the Haldane phase can also be found in or nearby the isotropic 2D limit on the triangular lattice, potentially offering further understanding of the unusual behavior of NiGa$_2$S$_4$~\cite{nakatsuji05} and Ba$_3$NiSb$_2$O$_9$~\cite{cheng11}.


\acknowledgments
We thank K. Harada and M. Matsumoto for providing us  Quantum Monte Carlo benchmark data and acknowledge insightful discussions with F. Mila, S.~Jiang, and T. Okubo. This project has received funding from the European Research Council (ERC) under the European UnionÕs Horizon 2020 research and innovation programme (grant agreement No 677061). This work is part of the \mbox{Delta-ITP} consortium, a program of the Netherlands Organization for Scientific Research (NWO) that is funded by the Dutch Ministry of Education, Culture and Science~(OCW). This research was supported in part by Perimeter Institute for Theoretical Physics. Research at Perimeter Institute is supported by the Government of Canada through Industry Canada and by the Province of Ontario through the Ministry of Research and Innovation.

\bibliographystyle{apsrev4-1}
\bibliography{../bib/refs,comments}

\begin{thebibliography}{73}
\bibitem [{COI()}]{COITMAA}%
  \BibitemOpen
  \href@noop {} {}\bibinfo {note} {In Ref.~\cite{oitmaa13} a consistent value
  of $\theta_c=0.190\pi$ was predicted for the vanishing of the AF order based
  on fitted series expansion data, from which the authors concluded that there
  is a continuous or a very weak first order transition between the AF and the
  3-sublattice phase.}\BibitemShut {Stop}%
\bibitem [{CVB()}]{CVBC}%
  \BibitemOpen
  \href@noop {} {}\bibinfo {note} {We have also checked for potential
  valence-bond solid states in unit cells up to $4\times4$ but could not find
  stable solutions.}\BibitemShut {Stop}%
\bibitem [{CU()}]{CU}%
  \BibitemOpen
  \href@noop {} {}\bibinfo {note} {A special full-update imaginary-time
  evolution algorithm has been used in this case which will be discussed
  elsewhere.}\BibitemShut {Stop}%
\end{thebibliography}

\end{document}


\title{Emergent Haldane phase in the $S=1$ bilinear-biquadratic Heisenberg model on the  square lattice: \textit{supplemental material} }

\author{Ido Niesen}
\affiliation{Institute for Theoretical Physics and Delta Institute for Theoretical Physics, University of Amsterdam, Science Park 904, 1098 XH Amsterdam, The Netherlands}

\author{Philippe Corboz}
\affiliation{Institute for Theoretical Physics and Delta Institute for Theoretical Physics, University of Amsterdam, Science Park 904, 1098 XH Amsterdam, The Netherlands}

\date{May 2, 2017}

\maketitle

\section{Detailed data of the iPEPS simulations of the anisotropic model}
\begin{figure}[b]
\includegraphics[width=0.6\columnwidth]{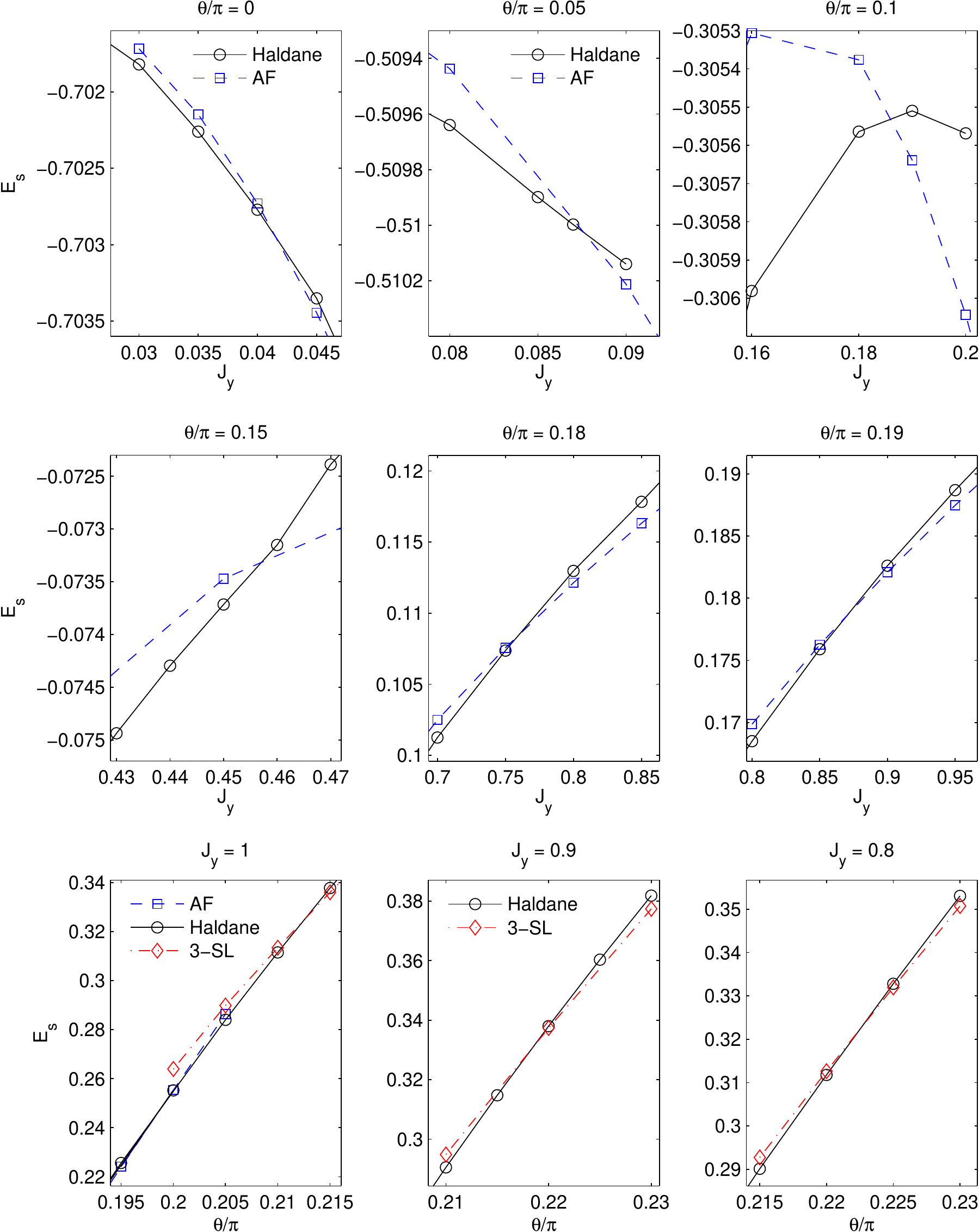} 
\caption{iPEPS energies for $D=10$ (simple update) as a function of $J_y$ (or $\theta/\pi$) for fixed values of $\theta/\pi$ (or $J_y$).
}
\label{fig:res_bilbiq_aniso3ipaper}
\end{figure}

In Fig.~\ref{fig:res_bilbiq_aniso3ipaper} we provide the iPEPS energies (for $D=10$, simple update) for different cuts in the phase diagram of the anisotropic model which were used to get an estimate of the phase boundary between the Haldane and AF phase, or Haldane and 3-sublattice phase, respectively (see main text). The phase transitions occur where the energies of the states intersect. This data was used to plot the phase diagram of the anisotropic model shown in Fig. 3 in the main text.

